\documentclass[reprint,a4paper,twocolumn,aps,prb,nopacs,superscriptaddress,longbibliography]{revtex4-2}
\usepackage{float}
\usepackage{soul}
\usepackage[english]{babel}
\usepackage[utf8]{inputenc}
\usepackage{fancyhdr}
\usepackage{sidecap}
\usepackage{multirow}

\usepackage{ulem}
\usepackage{graphicx}
\usepackage{mhchem}
\usepackage[utf8]{inputenc}
\usepackage{newunicodechar}
\newunicodechar{μ}{\ensuremath{\mu}}
\usepackage{braket}
\usepackage{tikz}
\usepackage{color}
\usepackage{amsmath}
\usepackage[colorlinks=True, linkcolor=blue, filecolor=magenta, urlcolor=blue,citecolor=blue]{hyperref}
\usepackage[nameinlink,capitalise]{cleveref}
\usepackage{comment}
\begin{document}

\title{Substitution effects in RuO$_2$ single crystals}

\author{Shubhankar Paul}
  \email{shubhp@iitk.ac.in}
\affiliation{Toyota Riken–Kyoto University Research Center (TRiKUC), Kyoto 606-8501, Japan}
\affiliation{Department of Electronic Science and Engineering, Graduate School of Engineering, Kyoto University, Kyoto 615-8510, Japan}
\affiliation{Department of Physics, Indian Institute of Technology Kanpur, Kanpur 208016, India}
\author{Kunihiko Yamauchi}
\affiliation{Center for Spintronics Research Network (CSRN), The University of Osaka, Toyonaka, Osaka 560-8531, Japan }
\author{Shogo Yamashita}
\affiliation{Max-Planck-Institute for Chemical Physics of Solids (MPI-CPfS), Dresden 01187, Germany}
\author{Hisakazu Matsuki}
\affiliation{Toyota Riken–Kyoto University Research Center (TRiKUC), Kyoto 606-8501, Japan}
\affiliation{Institute for Chemical Research, Kyoto University, Uji, Kyoto 611-0011, Japan}
\affiliation{Center for Spintronics Research Network (CSRN), Kyoto University, Uji, Kyoto 611-0011, Japan.}
\author{Shuhei Iwashita}
\affiliation{Division of Chemistry, Graduate School of Science, Kyoto University, Kyoto 615-8502, Japan}
\author{Mitsuhiko Maesato}
\affiliation{Division of Chemistry, Graduate School of Science, Kyoto University, Kyoto 615-8502, Japan}
\author{Hiroshi Kitagawa}
\affiliation{Division of Chemistry, Graduate School of Science, Kyoto University, Kyoto 615-8502, Japan}
\author{Chanchal Sow}
\affiliation{Department of Physics, Indian Institute of Technology Kanpur, Kanpur 208016, India}
\author{Shingo Yonezawa}
\affiliation{Department of Electronic Science and Engineering, Graduate School of Engineering, Kyoto University, Kyoto 615-8510, Japan}
\author{Yoshiteru Maeno}
\email{maeno.yoshiteru.b04@kyoto-u.jp}
\affiliation{Toyota Riken–Kyoto University Research Center (TRiKUC), Kyoto 606-8501, Japan}
\affiliation{Max-Planck-Institute for Chemical Physics of Solids (MPI-CPfS), Dresden 01187, Germany}
\affiliation{Faculty of Engineering, Kyoto University of Advanced Science (KUAS), Kyoto 615-8577, Japan}
\date{\today}
\begin{abstract}
RuO$_2$ has emerged as one of the leading candidates for investigating altermagnetism. Recent quantum oscillation and angle-resolved photoemission spectroscopy measurements found the absence of altermagnetism or antiferromagnetism in pure RuO$_2$ crystals. The continuing debate over intrinsic magnetic order in RuO$_2$ may reflect its proximity to an altermagnetic instability.
In this work, we grow single crystals of Ru$_{0.9}$V$_{0.1}$O$_2$ and investigate their structural, transport, and magnetic properties.
X-ray photoelectron spectroscopy reveals an average V oxidation state near $+4$.
The paramagnetic susceptibility remains nearly unchanged up to room temperature, with no evidence of magnetic ordering.
Thus, the 10\% V-substitution in RuO$_2$ does not induce altermagnetism.
Electronic structure calculations for the V-substituted systems using two methods suggest that a higher level of V-substitution leads to a significant change in the density of states.
These findings underscore the potential of nonmagnetic substitution in RuO$_2$ as an attractive candidate for probing altermagnetic transitions and their experimental signatures.

\end{abstract}
\maketitle

\section{Introduction}
The discovery of altermagnetism has introduced a new framework for understanding collinear magnetic materials beyond the conventional ferromagnets (FMs) and antiferromagnets (AFMs). 
In altermagnets (AMs), crystal symmetry enforces zero net magnetization while breaking time-reversal symmetry, giving rise to spin-dependent band structures and novel phenomena such as anomalous Hall effect \cite{naka2019spin,hayami2019momentum,vsmejkal2020crystal,vsmejkal2022beyond}, anisotropic spin splitting \cite{vsmejkal2022emerging}, spin currents \cite{karube2022observation, bose2022tilted}, and characteristic magnon excitations \cite{vsmejkal2022beyond}.
Among the proposed altermagnetic candidates, RuO$_2$ has attracted particular attention due to the prediction of large spin splitting with the order of 1.4 eV \cite{vsmejkal2022emerging}.
Rutile-structured RuO$_2$ is a well-established metallic oxide widely used as an oxygen evolution reaction (OER) catalyst \cite{Ping2024catalyst,Over2000catalyst} and a low-temperature thermometer \cite{bat1995design}. In addition, superconductivity has been realized in RuO$_2$ thin films under large uniaxial strain \cite{uchida2020superconductivity,ruf2021strain}.

RuO$_2$ was long regarded as a Pauli paramagnetic metal, but neutron diffraction measurements later suggested that its ground state is antiferromagnetic with a small ordered moment of 0.05~$\mu_{\mathrm{B}}$ per Ru atom \cite{berlijn2017itinerant}.
Such antiferromagnetic order with a Néel temperature $T_\mathrm{N} > 300$ K was confirmed by resonant X-ray scattering \cite{zhu2019anomalous} and ARPES \cite{Lin2024ARPES}.
Several experimental investigations of RuO$_2$ thin films have revealed altermagnetic behavior \cite{guo2024direct,he2025evidence}.
Recent studies on high-quality RuO$_2$ bulk single crystals show that pristine RuO$_2$ is paramagnetic rather than antiferromagnetic or altermagnetic \cite{hiraishi2024nonmagnetic,wu2025fermi,osumi2025ARPES,paul2025nonanalytic,kessler2024absence}.
However, theoretical calculations indicate that RuO$_2$ is close to altermagnetic instability \cite{smolyanyuk2024fragility,smolyanyuk2025origin}.
Thus, the magnetic ground state of RuO$_2$ films remains controversial.
An important question is whether altermagnetism in RuO$_2$ can be stabilized by partially substituting Ru with hole- or electron-donor dopants.
Recent theoretical studies suggest that hole doping and epitaxial strain stabilize the altermagnetic state \cite{qian2025fragile,smolyanyuk2024fragility,qian2025fragile}, potentially explaining the discrepancies among experimental observations.
In this regard, Cr-substituted RuO$_2$ thin films \cite{Wang-Kagawa2023NatComRuCrO2} have recently been reported to exhibit antiferromagnetic order.
However, subsequent studies by density functional theory (DFT) calculations \cite{smolyanyuk2025origin} demonstrated that the observed magnetism originates from the magnetic moments of the Cr dopants rather than from intrinsic RuO$_2$.
Although Nb and V doping in RuO$_2$ has been studied for developing high-performance acidic OER catalysts \cite{wu2025engineering,chun2013highly,liu2026recent}, its impact on altermagnetism remains largely unexplored.
Therefore, partial substitution with nonmagnetic elements is highly desirable to resolve this controversy and clarify the possible emergence of altermagnetism in RuO$_2$.

In this paper, we report the single-crystal growth of RuO$_2$ partially substituted with nonmagnetic elements and investigate its physical properties.
Among the candidate substituents, only vanadium is homogeneously incorporated into the RuO$_2$ lattice.
The average vanadium valence state is close to +4.
The room-temperature resistivity decreases by $\sim60\%$ upon V-substitution of 10\%.
Such a decrease is primarily attributed to the significant reduction in the electron-phonon scattering strength.
The paramagnetic susceptibility exhibits only a slight change up to room temperature without any sign of magnetic ordering.
Electronic structure calculations using two different methods indicate that V-$d$ states appear within 0.3 eV above the Fermi energy by V substitution. 

\begin{figure*}
	\begin{center}
		\includegraphics[scale=0.89]{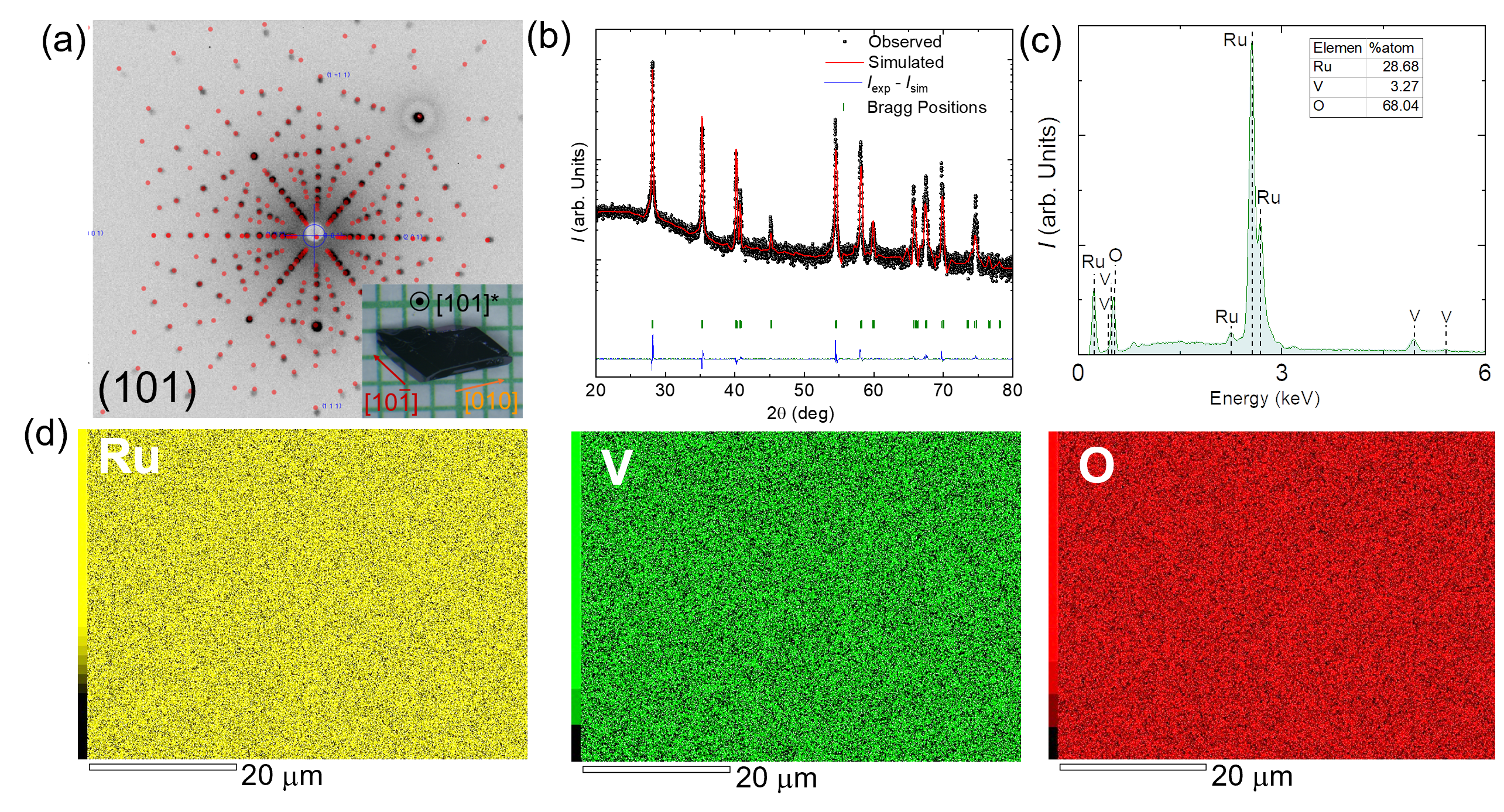}
		\caption{Structural characterization of single crystalline Ru$_{1-x}$V$_x$O$_{2+y}$. (a) X-ray Laue diffraction pattern on the (101) surface (black) along with simulated pattern (red). Inset shows the optical image of a typical crystal. (b) Powder XRD spectrum of a crushed crystal, showing no impurity peaks. The result of Rietveld refinement is shown in red. (c) EDS spectra of a grown crystal indicating the stoichiometric ratio of Ru$_{1-x}$V$_x$O$_{2+y}$. (d-f) Elemental mappings of Ru-L (left), V-K (center), and O-K (right) obtained by EPMA, suggesting uniform distribution of constituent elements.}
		\label{fig1}
	\end{center}
\end{figure*}

\section{Experimental}
In the present work, various elements were investigated as potential substituents for Ru in RuO$_2$ as summarized in Table \ref{tab1}.
The phase purity of the resulting  Ru$_{1-x}X_x$O$_{2+y}$ compounds was examined, and among the investigated elements, only vanadium formed a single-phase polycrystalline compound, Ru$_{0.9}$V$_{0.1}$O$_{2+y}$, which was subsequently selected for single-crystal growth.
Single crystals of Ru$_{0.9}$V$_{0.1}$O$_{2}$ were grown by the sublimation transport method \cite{paul2025growth} using RuO$_2$ (Rare Metallic, 99.95\%) and VO$_2$ (High Purity Chemicals, 99.9\%) as starting materials.
First, RuO$_2$ and VO$_2$ powders were mixed in a 1:1 molar ratio and pressed into a pellet following the procedure described in~\cite{paul2025growth}.
The pellet was then placed in the hot zone of a necked-shaped growth tube and heated to 1000$^\circ$C for 20 h.
Subsequently, the temperature was increased to 1200$^\circ$C and maintained for 100 h with an O$_2$ gas (99.999\%) flow rate of 40 cc/min.
Single crystals were obtained in the colder region of the growth tube through sublimation vapor transport.

Phase purity of the crystals were verified by X-ray diffraction (XRD) using a Rigaku Miniflex 600-C diffractometer with a Cu-K$_\alpha$ source.
Elemental compositions of the grown crystals were determined using an electron probe microanalyzer (EPMA) (JEOL, JXA-8230) through energy-dispersive X-ray spectroscopy (EDS) and wavelength-dispersive X-ray spectroscopy (WDS) measurements.
The oxidation states of Ru, V, and O were investigated by X-ray photoelectron spectroscopy (XPS) (SHIMADZU, ESCA-3400) equipped with a monochromatic Al-K$_\alpha$ X-ray source (1486.6 eV).
The base pressure during the measurements was maintained at the order of 10$^{-7}$~Pa.
The XPS spectra were analyzed and fitted using XPS PeakFit 4.1 software \cite{kwok2000xps}.
All binding energies (B.E.) were corrected with respect to the adventitious carbon C 1s peak at 284.8 eV.
A full survey energy scan was also performed within 1-20 minutes to inspect the elements present in the system.

Electrical resistivity was measured using a custom-built transport probe mounted in a MPMS-XL system (Quantum Design).
Standard four-probe AC measurements were performed with a lock-in amplifier (Stanford Research Systems, SR830) operating at a frequency of 17 Hz and an excitation current of 10 mA (rms). Magnetization measurements were carried out using a superconducting quantum interference device magnetometer (Quantum Design, MPMS-XL).

\begin{figure*}
	\begin{center}		\includegraphics[scale=0.70]{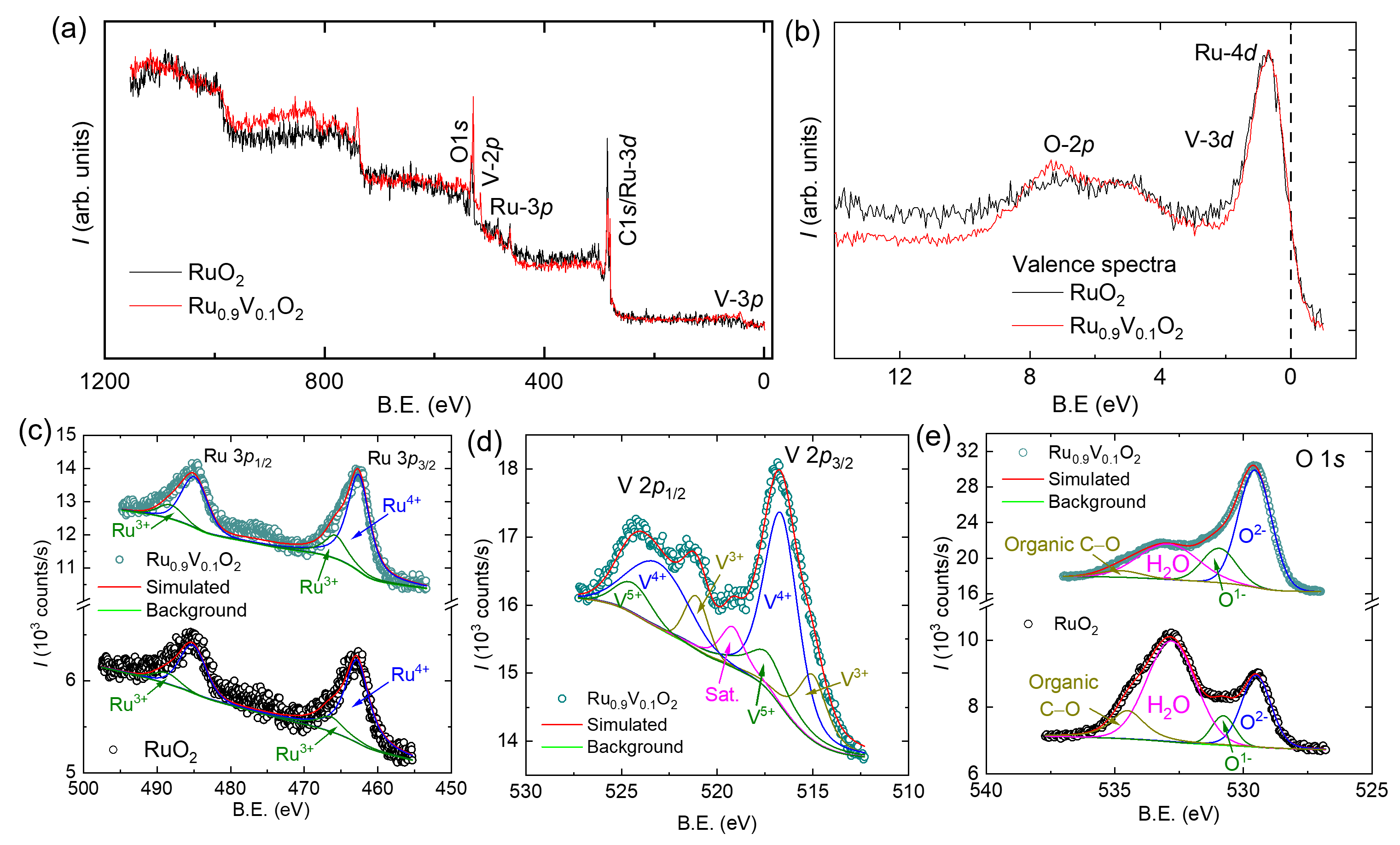}
		\caption{XPS spectra of RuO$_{2+y}$ and Ru$_{1-x}$V$_x$O$_{2+y}$ single crystals. (a) Survey spectra of both crystals, indicating the absence of impurity elements. (b) The valence spectra of both crystals. (c) Comparison of the core level spectra of Ru-3$p$ for both samples. In both samples, Ru$^{4+}$ as well as a slight amount of Ru$^{3+}$ is detected. (d) Core level spectrum of V-2$p$ for V-substituted RuO$_2$ crystal. It is dominated by V$^{4+}$ contribution, but V$^{5+}$ as well as V$^{3+}$ contributions are also visible. (e) Comparison of O-1$s$ core level spectra of both crystals. The peak for O$^{1-}$ is attributed to surface hydroxide.}
		\label{fig2}
	\end{center}
\end{figure*}

\begin{table*}
    \setlength{\tabcolsep}{1.5pt}    
    \fontsize{8}{10}\selectfont
 	\caption{Syntheses conditions to introduce potential substituent elements for Ru in Ru$_{1-x}X_x$O$_{2+y}$.
    The first group represents polycrystalline synthesis attempting substitution by potentially non-magnetic elements using a box furnace.
    The second group shows the results of single-crystal growth using a tube furnace.\vspace{2pt}
    }
 	\label{tab1}
 	\begin{tabular*}{1.0\textwidth}{@{\extracolsep{\fill}}ccccccccccl}
 		\hline
 		\hline
 		Substituents&Starting powders&Nominal&Atmosphere&Temperature &Time&X-ray&Lattice\\
      ($X$)&for substituent&$x$  &  &($^\circ$C)&(hr)&peaks&parameters ($\mathring{\text{A}}$)\\
		\hline \vspace{2pt}
        Polycrystalline synthesis &&&&&&&\\
    	  Ti & Ti$_2$O$_3$  & 0.05 & air & 1000 & 20 &RuO$_2$, TiO$_2$ &*&  \\
        Nb & NbO$_2$  & 0.05 & air & 1000 & 20 &RuO$_2$, NbO$_2$&*&  \\
        Zr & ZrO$_2$  & 0.05 & air & 1000, 1100, 1200 & 20 &RuO$_2$, ZrO$_2$&*&  \\
        Zr & ZrO$_2$  & 0.1 & air & 1000 & 20 &RuO$_2$, ZrO$_2$ &*&  \\
        Ta & Ta$_2$O$_5$  & 0.05 & air & 1000, 1100, 1200 & 20 &RuO$_2$, Ta$_2$O$_5$&*&  \\
        Sn& SnO  & 0.1 & air & 1000 & 20 &RuO$_2$, SnO, SnO$_2$&*&  \\
        Ge & GeO$_2$  & 0.1 & air & 1000, 1100, 1200 & 20 &RuO$_2$, GeO$_2$ &*&  \\
        Al & Al$_2$O$_3$  & 0.1 & air & 1000 & 20 &RuO$_2$, Al$_2$O$_3$ &*&  \\
        Sc & Sc$_2$O$_3$  & 0.05 & air & 1000 & 20 &RuO$_2$, Sc$_2$O$_3$ &*&  \\
        V & V$_2$O$_3$  & 0.1 & air & 1000 & 24 &RuO$_2$, V$_2$O$_3$ &*&  \\
        V& V$_2$O$_5$  & 0.1 & air & 1000 & 24 &RuO$_2$, V$_2$O$_5$ &*&  \\ \vspace{2pt}
       Single-crystal growths &&&&&&&\\
        RuO$_{2-y}$& RuO$_2$  & 0 &95\%Ar+5\%H$_2$& 600 & 20 & Ru metal &*&  \\
        \vspace{0.1 mm}
        & & &flowing &&  & only & metal&  \\
        Ti & Ti$_2$O$_3$  & 0.10 & O$_2$ flowing &1000(20 hr)+1250(85 hr)& 105 &RuO$_2$, TiO$_2$&*&  \\
        V & VO$_2$  & 0.1 & O$_2$ flowing &1000(20 hr)+1250(100 hr)& 124 & RuO$_2$ &$a=4.496$, \\
        & & & && & only & $c=3.092$&  \\        
         \hline
         \hline
	\end{tabular*}
\end{table*}

\section{Results and discussion}
\subsection{Crystal purity and homogeneity: XRD and EPMA}

Figure~\ref{fig1}\textcolor{blue}{(a)} displays an optical image and a back-scattered Laue diffraction photo of Ru$_{1-x}$V$_{x}$O$_{2+y}$ single crystals exhibiting a flat-like morphology, with typical dimensions of approximately  1.5 $\times$ 2 $\times$ 5 mm$^3$.
The phase purity and lattice parameters of the crystals are investigated by powder XRD using finely ground crystals.
The diffraction pattern, collected over a wide angular range down to 2$\theta$ = 3$^\circ$, exhibits only reflections attributable to the RuO$_2$ phase, indicating the absence of detectable impurity phases, as shown in Fig.~\ref{fig1}\textcolor{blue}{(b)}.
Rietveld refinement of the powder XRD data using the FullProf suite \cite{rodriguez1993recent} gives lattice parameters of $a~=~4.496$ $\mathring{\text{A}}$ and $c~=~3.092$~$\mathring{\text{A}}$.
Compared to pristine RuO$_2$ \cite{paul2025growth}, the $a$-axis exhibits a slight expansion of 0.12\%, whereas the $c$-axis contracts by 0.58\%, leading to an overall unit cell volume reduction of $0.38\%$, consistent with the smaller ionic radius of V$^{4+}$ compared to Ru$^{4+}$.
A similar trend in lattice parameters with increasing V content has been reported for Ru$_{1-x}$V$_{x}$O$_{2}$ nanowires \cite{doi:10.1021/am4016445}, showing a linear variation consistent with Vegard's rule for a continuous substitutional solid solution \cite{wang2005structural,doi:10.1021/am4016445}.

To examine the elemental composition and spatial distribution of the constituent elements, EDS and WDS measurements were carried out on both the as-grown surface and a freshly polished surface as shown in Figs.~\ref{fig1}\textcolor{blue}{(c)} and \textcolor{blue}{(d)}.
The elemental composition of the crystals, determined by WDS [Fig.~\ref{fig1}\textcolor{blue}{(c)}], is close to the nominal stoichiometric ratio, yielding Ru~:~V~:~O~=~ 0.89~:~0.10~:~2.37.
Although the measurements were carried out under high-vacuum conditions, the slightly higher oxygen content is likely attributable to a small amount of oxygen adsorbed on the crystal surface, which can contribute to the WDS signal.
Considering this surface-related effect and the inherent uncertainty in oxygen quantification by a WDS detector, the crystal composition is assumed to be Ru$_{0.9}$V$_{0.1}$O$_{2+y}$.
Figures \ref{fig1}\textcolor{blue}{(d)}-\textcolor{blue}{(f)} show elemental mapping with high spatial resolution across the surface area of 45~$\times$~60 mm$^2$ of a polished crystal.
In each elemental mapping, bright and dark regions correspond to element-rich and element-poor areas, respectively.
These elemental mappings reveal uniform distributions of Ru, V, and O throughout the examined region, indicating excellent compositional homogeneity without clustering of~vanadium.

\subsection{Oxidation states: XPS}
To examine the oxidation state of the constituent element in Ru$_{0.9}$V$_{0.1}$O$_{2+y}$ crystal the XPS analysis was performed.
Figure~\ref{fig2}\textcolor{blue}{(a)} shows the XPS survey spectra of RuO$_2$ and Ru$_{0.9}$V$_{0.1}$O$_2$ crystals, revealing only the peaks of Ru-3$p$, 3$d$, V-2$p$, 3$p$, O-1$s$, and adventitious C-1$s$, confirming the absence of impurity elements.
The C-1$s$ peak originates from atmospheric hydrocarbon contamination.
Figure~\ref{fig2}\textcolor{blue}{(b)} presents the room-temperature valence band (VB) spectra of both samples.
The spectra are dominated by Ru 4$d$ states near the Fermi level ($\epsilon_\mathrm{F}$, 0–2 eV), whereas the O-2$p$ states span the 2.35–11 eV energy range.

Figure~\ref{fig2}\textcolor{blue}{(c)} presents the Ru~3$p$ core-level XPS spectra of RuO$_2$ and Ru$_{0.9}$V$_{0.1}$O$_2$.
It exhibits a characteristic Ru~3$p$ spin--orbit doublet with the 3$p_{3/2}$ and 3$p_{1/2}$ peaks at 462.8 and 485.0~eV, respectively, separated by 22.2~eV.
The Ru 3$p$ core-level spectra of RuO$_2$ and V-doped RuO$_2$ exhibit mixed Ru$^{4+}$/Ru$^{3+}$ valence states with relative area fractions of 80\% and 20\%, respectively, while no metallic Ru is observed in both crystals.
The Ru$^{3+}$ inclusion is essential to account for the shoulder feature near the main peak.
The presence of Ru$^{3+}$ state may be associated with surface reduction, although its spectral weight is considerably smaller than that reported previously \cite{wu2025engineering}.
The Ru$^{4+}$ binding energy remains essentially unchanged upon 10\% V doping.
The average valence state of Ru ions is determined by analyzing the fitted Ru $3p_{3/2}$ and $3p_{1/2}$ spectra with a fixed intensity ratio of $2:1$.
The fitted parameters are summarized in Table~\ref{tb1b}.
The resulting average Ru valence is $+3.83$ and $+3.80$ for RuO$_2$ and Ru$_{0.9}$V$_{0.1}$O$_2$, respectively.

Due to spin-orbit coupling, the V-2$p$ XPS spectrum exhibits two main peaks corresponding to V-2$p_{3/2}$ and V-2$p_{1/2}$ at binding energies of 516.6 and 523.0 eV, respectively.
The V-2$p$ core-level spectrum is deconvoluted into three oxidation-state, namely V$^{3+}$, V$^{4+}$, and V$^{5+}$ [Fig.~\ref{fig2}\textcolor{blue}{(d)}].
The V$^{3+}$ component is essential to explain the additional peak at 521 eV corresponding to V-$2p_{1/2}$.
An earlier study, which did not consider the V$^{3+}$ component, reported the spectra only up to 520 eV \cite{wu2025engineering}.
Among the three oxidation states, V$^{4+}$ is the dominant, indicating that vanadium is predominantly incorporated in the +4 oxidation state, while smaller fractions of V$^{3+}$ and V$^{5+}$  are also present.
The average oxidation state of V is estimated to be $+3.97$.

\begin{figure*}
	\begin{center}
		\includegraphics[scale=0.70]{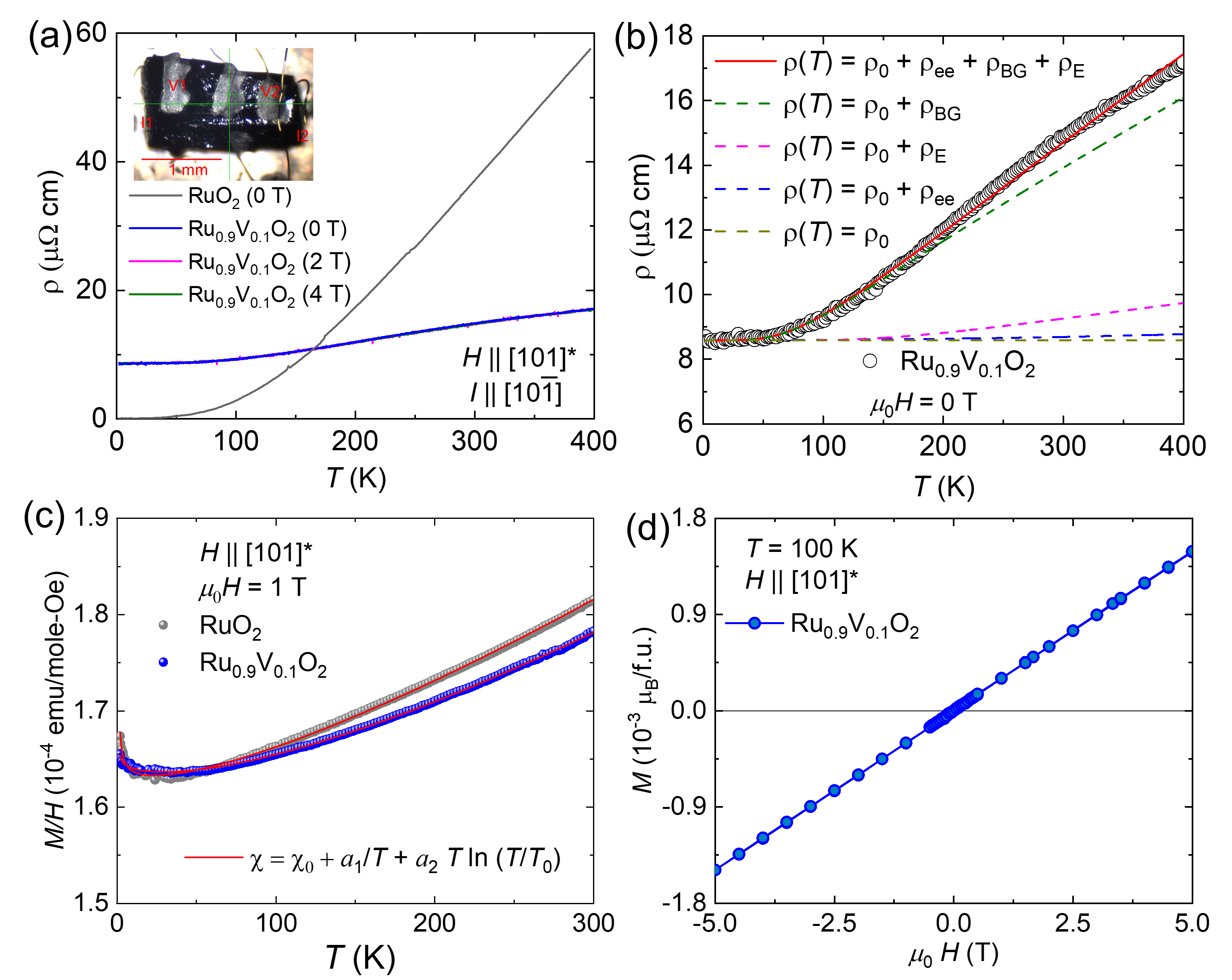}
		\caption{(a) Comparison of the temperature dependence of resistivity of RuO$_2$ \cite{paul2025growth} and V-substituted RuO$_2$ crystals. 
        The magnetic field dependence is also shown for the V-substituted RuO$_2$. 
        The top-left inset shows an optical image of the Ru$_{0.9}$V$_{0.1}$O$_2$ crystal with current applied along the [10$\bar{1}$] direction and field perpendicular to the top surface. 
        (b) The resistivity fit using Eqs.~\ref{RuO2eq5_1} and \ref{RuO2eq5_2} (red solid line). The yellow, blue, magenta, and green dashed lines represent impurity ($\rho_0$), electron–electron scattering ($\rho_\mathrm{ee}$), optical-phonon ($\rho_\mathrm{E}$) and acoustic-phonon ($\rho_\mathrm{BG}$) contributions to the total resistivity, respectively.
        (c) DC magnetic susceptibility ($\chi=M/H$) with magnetic field along the [101]* direction for both crystals. 
        The red curves represent the fitting with Eq.~\ref{RuO2_eq5_3}. 
        (d) Field-dependent magnetization of Ru$_{0.9}$V$_{0.1}$O$_2$ with the magnetic field applied along the [101]* direction at 100 K.}
		\label{fig3}
	\end{center}
\end{figure*}

The O-1s spectrum was deconvoluted into four components centered at 529.5, 530.7, 532.8 eV, and 534.5 eV [Fig.~\ref{fig2}\textcolor{blue}{(e)}].
The peak centered at 529.5 eV is attributed to anhydrous RuO$_2$ and corresponds to lattice oxygen (O$^{2-}$) within the rutile crystal structure.
The intermediate peak at 530.9 eV is attributed to O$^{1-}$ in hydroxide (OH$^-$) on the sample surface \cite{morgan2015resolving,wu2025engineering}.
The peak at 532.8 eV is attributed to loosely bound oxygen species, primarily from H$_2$O molecules on the surface.
The high-binding-energy peak at 534.5 eV corresponds to adsorbed oxygen-containing surface species (denoted as O$^\mathrm{Chem}$), including CO and CO$_2$, which may originate from surface contamination upon exposure to the ambient atmosphere.
This interpretation leads to the oxidation state of oxygen in the lattice to be $-2$ for RuO$_2$ and Ru$_{0.9}$V$_{0.1}$O$_2$.

The chemical composition of the samples can be represented as Ru$_{1-x}$V$_x$O$_{2+y}$, where $y$ is the oxygen off-stoichiometry per formula unit.
From the charge neutrality condition, the oxygen off-stoichiometry is estimated to be $-0.08$ for RuO$_{2+y}$ and $-0.09$ for Ru$_{0.9}$V$_{0.1}$O$_{2+y}$, suggesting that V-doping has a negligible effect on the oxygen stoichiometry.

\subsection{Resistivity and Magnetization}
Figure \ref{fig3}\textcolor{blue}{(a)} shows the temperature dependence of the resistivity of Ru$_{0.9}$V$_{0.1}$O$_2$ measured between 2 and 400~K under various magnetic fields, with the current applied along the [10$\bar{1}$] and the field along the [101]* directions.
Current leads were attached to the side surfaces of both ends of the crystal, and voltage leads to its top surface [inset of Fig.~\ref{fig3}\textcolor{blue}{(a)}] using silver epoxy (EPO-TEK H20E).
To minimize the contact resistance and obtain robust electrical contacts, the silver epoxy was annealed at 500$^\circ$C in an Ar atmosphere to remove its organic contents \cite{paul2025growth,paul2026CrSb,paul2026multi}. Annealing in an Ar environment prevents silver oxidation and minimizes contact resistance.
The Ru$_{0.9}$V$_{0.1}$O$_2$ exhibits metallic behavior throughout the measured temperature range, with a residual resistivity ratio (RRR = $\rho_{300,\mathrm{K}}/\rho_{2,\mathrm{K}}$) of 1.8.
The residual resistivity of 8.6~$\mu\Omega$ cm corresponds to an estimated mean free path of 1.5~nm.
The room-temperature resistivity of V-doped RuO$_2$ is nearly half of bulk RuO$_2$, suggesting enhanced metallic conduction.
The resistivity remains essentially unchanged under applied magnetic fields up to 4 T.
It is noted that the room-$T$ resistivity is reduced by 60\% in the V-doped system.

We fit the resistivity of Ru$_{0.9}$V$_{0.1}$O$_2$ using four distinct contributions \cite{paul2025nonanalytic},
\begin{equation}
\begin{split}
\rho(T) =\;& \rho_0 + A T^2
+ B\lambda_{\mathrm{BG}}
\left(\frac{T}{\theta_\mathrm{D}}\right)^4
\int_{0}^{\theta_\mathrm{D}/T}
\frac{x^5}{4\sinh^2(x/2)}\,dx \\
&+ \frac{B\lambda_{\mathrm{E}}}{4}
\left[
\frac{\theta_\mathrm{E}/2T}
{\sinh(\theta_\mathrm{E}/2T)}
\right]^2,
\end{split}
\label{RuO2eq5_1}
\end{equation}
\begin{equation}
         B = \frac{32\ \pi^2k_\mathrm{B}T}{\hbar\ \Omega_\mathrm{p}^2},
    \label{RuO2eq5_2}
\end{equation}

\noindent
The four terms in Eq.~\ref{RuO2eq5_1} correspond to the residual resistivity due to impurities and defects, electron-electron (e-e) scattering, the Bloch-Gr\"{u}neisen (BG) contribution from acoustic phonons, and the Einstein contribution from optical phonons, respectively.
In Eq.~\ref{RuO2eq5_2}, $\Omega_\mathrm{p}$ represents the plasma frequency.
Fitting the resistivity data in the temperature range of $2-30$~K without the phonon scattering terms in Eq.~\ref{RuO2eq5_1}, yields a quadratic coefficient $A$.
The resistivity data between 2 and 400~K is analyzed using Eqs.~\ref{RuO2eq5_1} and \ref{RuO2eq5_2}, with $\lambda_\mathrm{BG}$, $\lambda_\mathrm{E}$, $\theta_\mathrm{D}$, and $\theta_\mathrm{E}$ taken as adjustable fitting parameters.
The plasma frequency, $\hbar\Omega_\mathrm{p}=3.16$~eV, is fixed to the experimentally reported value for RuO$_2$ \cite{wenzel2025fermi}.
The fitting parameters are summarized in Table~\ref{tb2}.

\begin{table}[b]
    \setlength{\tabcolsep}{6pt}   
\fontsize{8}{10}\selectfont
    \caption{Comparison of the resistivity fitting parameters for RuO$_2$ \cite{paul2025growth} and Ru$_{0.9}$V$_{0.1}$O$_{2}$.}
    \label{tb2}
    \centering
    \begin{tabular}{cccc} 
 	\hline
    \hline 
     Parameter&RuO$_2$ & Ru$_{0.9}$V$_{0.1}$O$_{2}$\\
    \hline
   RRR & 1200 & 1.8 &  \\
    $\rho_0$ ($\mu\Omega$~cm) & 0.03 & 8.64\\
       $A$ (n$\Omega$ cm/K$^2$) & 0.051~ & 0.0012 \\ 
      $\lambda_{\mathrm{BG}}$ & 0.15& 0.05 \\ 
     $\lambda_{\mathrm{E}}$  & 0.22 & 0.01\\ 
    $\theta_\mathrm{D}$ (K) & 409 & 453 \\
    $\theta_\mathrm{E}$ (K) & 834 & 824 \\ 
    \hline
    \hline
 \end{tabular}
\end{table} 

For RuO$_2$, the total resistivity at 300 K is 36.1 $\mu\Omega$ cm, of which the BG and Einstein phonon contributions are 16.7 and 14.7 $\mu\Omega$ cm, corresponding to 46\% and 41\% of the total resistivity, respectively.
The e-e scattering contribution accounts for 13\%, while the residual resistivity contributes only 0.1\%, reflecting the RRR of 1200.
Upon V-substitution in RuO$_2$, the room-temperature resistivity decreases to $14.7~\mu\Omega$ cm, while $\rho_0$ is enhanced to $8.6~\mu\Omega$ cm.
Figure~\ref{fig3}\textcolor{blue}{(b)} illustrates individual contributions of different terms in Eq.~\ref{RuO2eq5_1} in the fitting. 
The temperature-dependent contributions arise primarily from the acoustic phonon term, which contributes 36\% at room temperature, while the Einstein phonon term contributes 5\%.
The $AT^2$ contribution is comparatively small, accounting for only 1\% of the total resistivity.
Our analysis indicates that substantial decrease in electron-phonon scattering strength leads to the reduction of room-temperature resistivity.
It is noted that $\theta_\mathrm{D}$ increases by 11\% upon 10\% V-substitution.
A similar enhancement of $\theta_\mathrm{D}$ by 7.4\% from the specific heat has been reported for 4.5\% Nb-doped RuO$_2$~\cite{mertig1986specific}.

\begin{figure*}
	\begin{center}
		\includegraphics[scale=0.52]{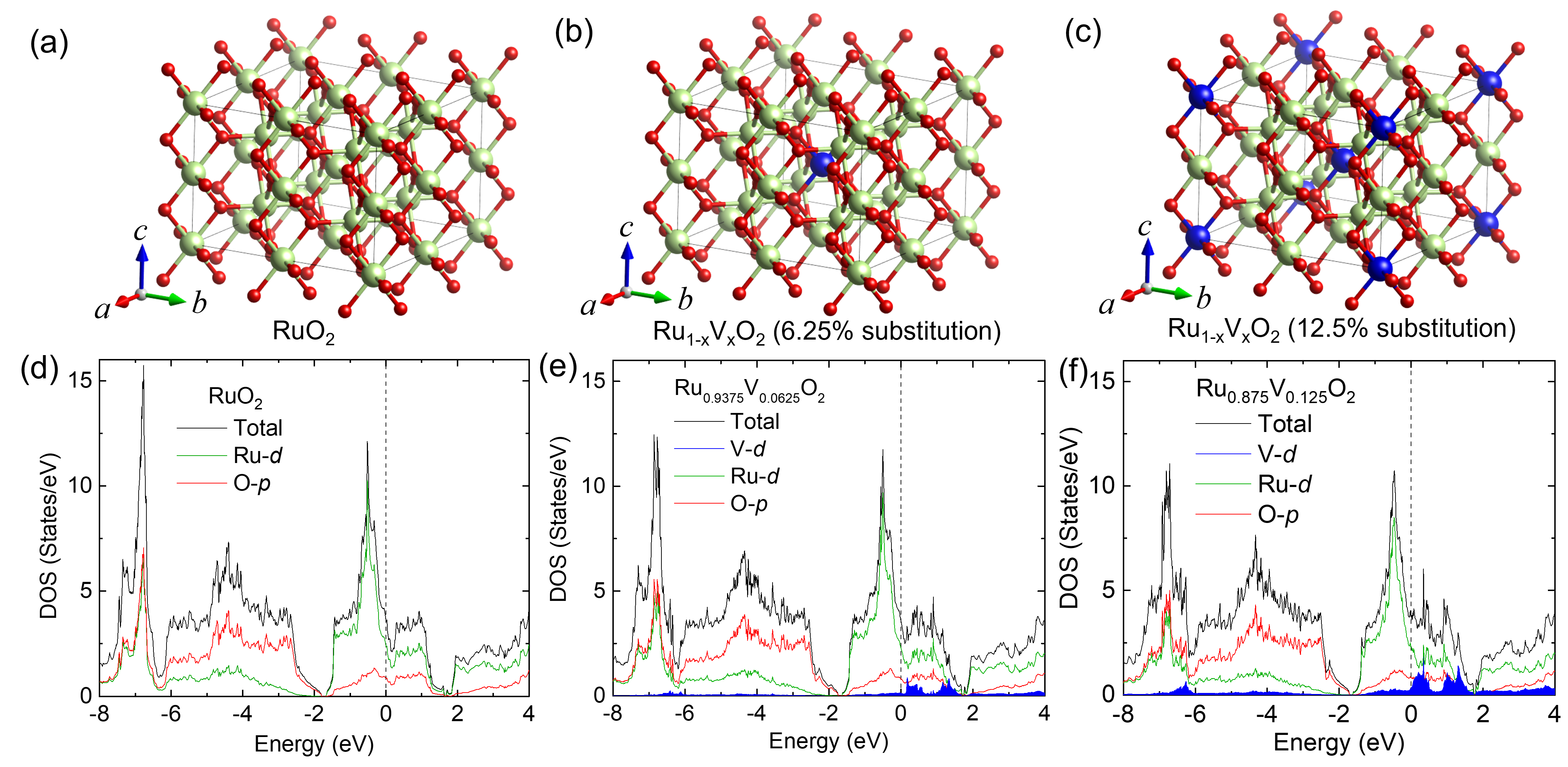}
		\caption{Comparison of the electronic density of states (DOS) calculated using the PAW method with GGA approximation. (a-c) Schematic representation of a $2\times2\times2$ supercell of RuO$_2$ and Ru$_{1-x}$V$_{x}$O$_2$. For 6.25\% V substitution, one V atom replaces one of the sixteen Ru sites, while for 12.5\%  substitution, two V atoms replace two Ru sites.
        The orbital-resolved DOS is presented for (d) RuO$_2$, (e) Ru$_{0.9375}$V$_{0.0625}$O$_2$, and (f) Ru$_{0.875}$V$_{0.125}$O$_2$. The DOS values are normalized per unit cell by dividing the original DOS values by eight. The total DOS is represented by the black line, while the projected DOS of Ru-$d$, O-$p$, and  V-$d$ orbitals are shown by the green line, red line, and blue shaded area, respectively.}
		\label{RuO2fig5_4}
	\end{center}
\end{figure*}

Figure~\ref{fig3}\textcolor{blue}{(c)} displays the dc susceptibility ($\chi=M/H$) of RuO$_2$ and Ru$_{0.9}$V$_{0.1}$O$_{2}$ single crystals measured with a magnetic field of 1~T applied along the [101]* direction.
Although $y$ is assumed to be zero in RuO$_{2+y}$ and Ru$_{0.9}$V$_{0.1}$O$_{2+y}$, this assumption does not affect the conclusions regarding the differences between the two systems.
The magnetic susceptibility decreases monotonically with decreasing temperature.
The slight upturn at low temperatures is attributed to the presence of paramagnetic impurities. 
V-substituted RuO$_2$ exhibits paramagnetic behavior with no magnetic transition between 1.8 and 300 K.
Interestingly, the low-temperature upturn is weaker in Ru$_{0.9}$V$_{0.1}$O$_2$ ($\chi_{1.8,\mathrm{K}}/\chi_{20,\mathrm{K}} = 1.01$) than in RuO$_2$ ($\chi_{1.8,\mathrm{K}}/\chi_{20,\mathrm{K}} = 1.02$), indicating a reduced paramagnetic impurity contribution upon V substitution.
However, V doping leads to a reduction in the room-temperature magnetic susceptibility.
The $\chi-T$ data is fitted using a logarithmic temperature-dependent term commonly used for $d$-electron transition metals \cite{paul2025nonanalytic}:
\begin{equation}
\begin{split}
    \chi \left(T,H\right) &= \chi_0 + \frac{a_1}{T} + a_2~T~\mathrm{ln}\left(\frac{T}{T_0}\right).
    \end{split}
    \label{RuO2_eq5_3}
\end{equation}
Where $\chi_0$ corresponds to Pauli susceptibility, $a_1/T$ accounts for the paramagnetic impurity contribution, and the logarithmic term is attributed to the intrinsic temperature dependence arising from the lattice expansion–induced variation of orbital paramagnetism.
From the fitting we obtain $\chi_0 = 1.63  \times 10^{-4}$ and $1.64 \times 10^{-4}$ emu/mol, characteristic temperature $T_0=30$ and 74 K and logarithmic term $a_2= 2.67 \times 10^{-8}$ and  $3.27 \times 10^{-8}$ emu/K$^2$~mol for RuO$_2$ and Ru$_{0.9}$V$_{0.1}$O$_2$, respectively.
Assuming that the paramagnetic impurities are localized spin-1 Ru$^{4+}$ ions with a concentration $z$, the Curie contribution is given by 
\begin{equation}
a_1=\frac{(g\mu_B)^2S(S+1)z}{3k_B}=0.5z.
\end{equation}
Fitting the susceptibility data using Eq.~(\ref{RuO2_eq5_3}) yields $a_1 = 9.5\times10^{-6}$ and $1.97\times10^{-6}$ emu K/mol, corresponding to impurity concentrations of $\sim$19 ppm for RuO$_2$ and $\sim$4 ppm for Ru$_{0.9}$V$_{0.1}$O$_2$. These results indicate that V doping reduces the concentration of localized paramagnetic impurities.
The field-dependent magnetization of  Ru$_{0.9}$V$_{0.1}$O$_2$ measured at 100 K is shown in Fig.~\ref{fig3}(d). The nearly linear, non-hysteretic nature is consistent with paramagnetic behavior \cite{paul2025growth}.

\subsection{Electronic Structure}
To examine the effect of V-doping on the electronic structure, we performed first-principles electronic-structure calculations using two distinct approaches: 1.~the electron projector augmented-wave (PAW) method and 2.~the tight-binding linear muffin-tin orbital (TB-LMTO) method.
The method 1 generally gives more accurate results for the electronic structure, while the method 2 is more suited to study chemical substitution or disordered systems.

\begin{figure}
	\begin{center}
		\includegraphics[scale=0.45]{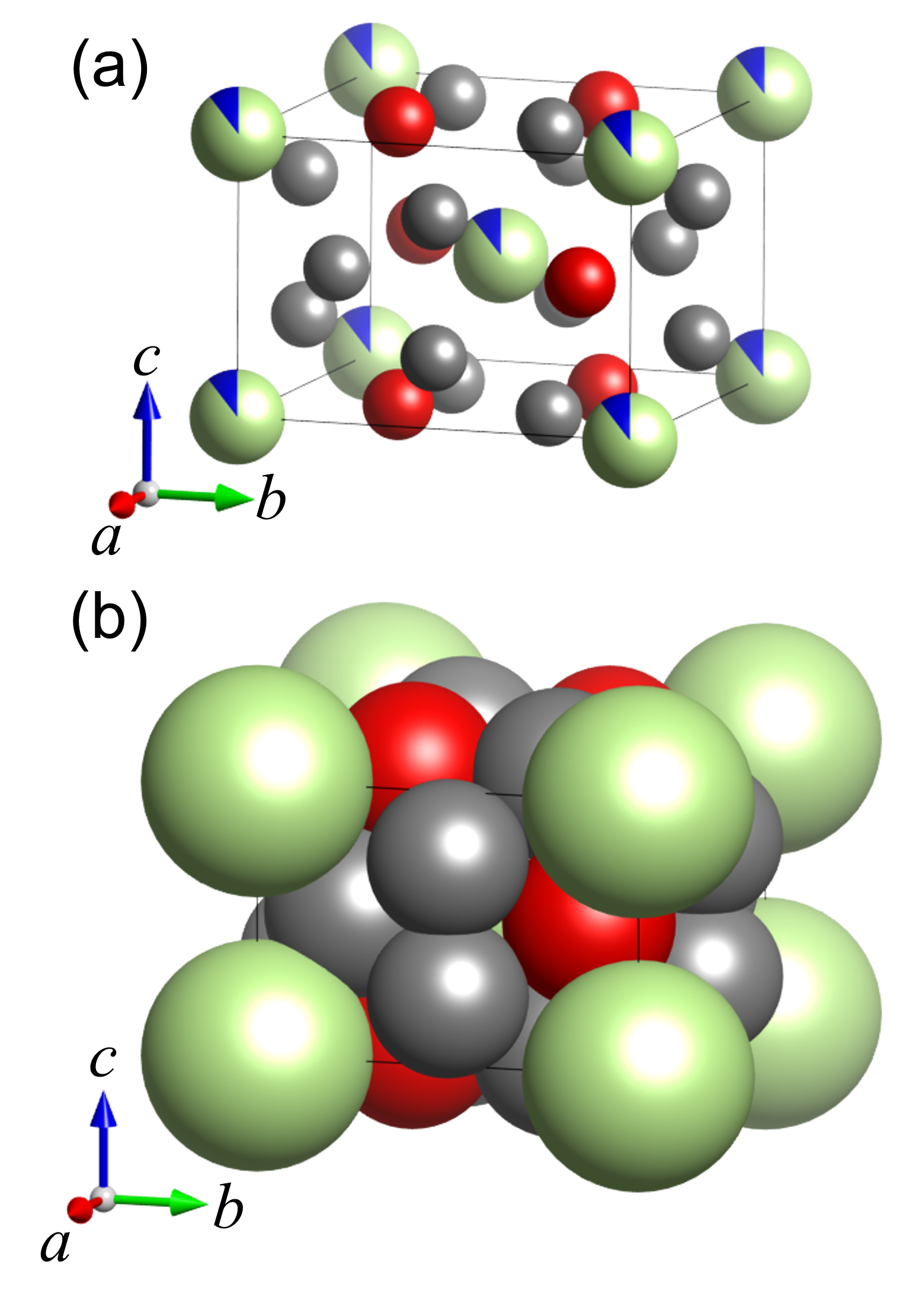}
		\caption{(a) Schematic representation of the crystal structure of Ru$_{0.9}$V$_{0.1}$O$_2$. Vanadium (blue) is considered for all Ru sites (green). The light gray spheres represent the locations of the empty spheres introduced solely for the ASA space-filling construction. (b) The cell volume filled with spheres with the radii given in Table~\ref{tb4}}
		\label{RuO2fig5_5}
	\end{center}
\end{figure}

\begin{figure*}
	\begin{center}
		\includegraphics[scale=0.65]{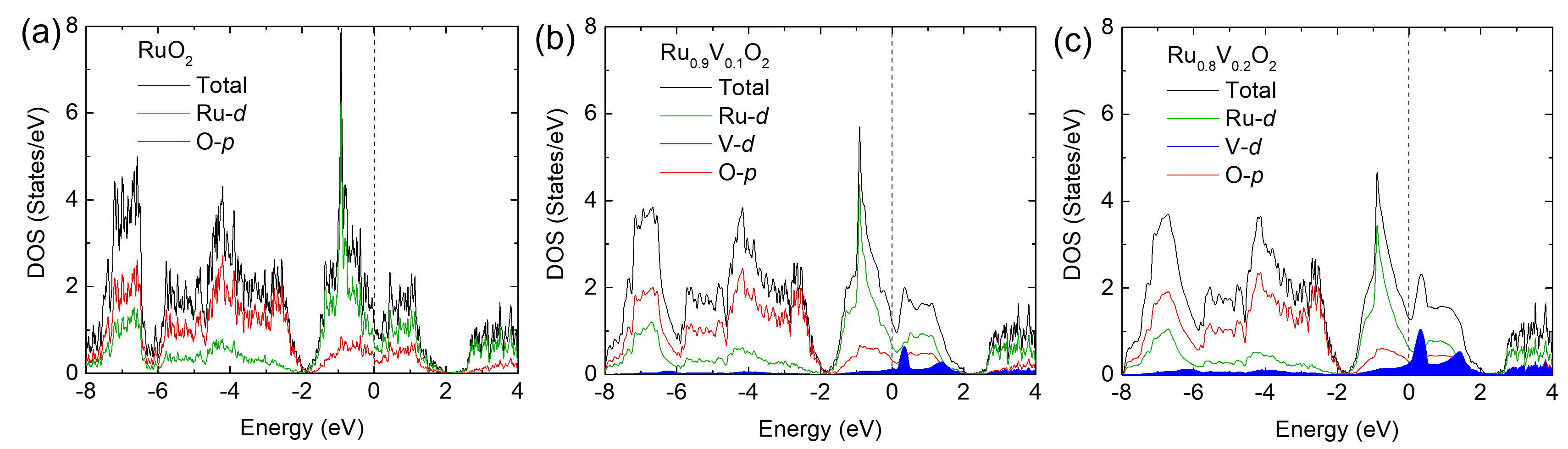}
		\caption{Comparison of the electronic density of states (DOS) calculated using the TB-LMTO-ASA+CPA method. The orbital-resolved DOS is presented for (a) RuO$_2$, (b) Ru$_{0.9}$V$_{0.1}$O$_2$, and (c) Ru$_{0.8}$V$_{0.2}$O$_2$. The total DOS is shown in black, while the projected DOS of Ru-$d$, O-$p$, and V-$d$ orbitals are shown by the green line, red line, and blue shaded area, respectively.}
		\label{RuO2fig6_6}
	\end{center}
\end{figure*}

\subsubsection{Projector augmented-wave (PAW) method}
The first approach is the PAW method based on the density functional theory (DFT), as implemented in the VIENNA \textit{Ab Initio} simulation package (VASP).
The PAW method provides an accurate description of the electronic structure and allows explicit modeling of chemical substitution using a supercell.
An advantage of this approach is that the local atomic environment around the substituted sites can be treated directly, including structural relaxation.

The exchange-correlation effects are considered within the generalized gradient approximation (GGA) using the Perdew-Burke-Ernzerhof (PBE) functional form without a Hubbard correction, corresponding to $U = 0$ in the GGA+$U$ framework.
Figure~\ref{RuO2fig5_4} represents the orbital-resolved DOS along with the corresponding supercell structures for RuO$_2$ and Ru$_{1-x}$V$_x$O$_2$.
The calculations were performed using a $2\times2\times2$ RuO$_2$ supercell containing 16 Ru sites, with experimental lattice parameters $a=4.4919$ $\mathring{\text{A}}$, $c=3.1066$ $\mathring{\text{A}}$ \cite{mattheiss1976electronic}, and internal atomic coordinates were relaxed until residual forces became smaller than 0.01 eV/$\mathring{\text{A}}$.

We consider three cases: pristine RuO$_2$ with no V substitution [Fig.~\ref{RuO2fig5_4}\textcolor{blue}{(a)}], one V atom substituting for one of the sixteen Ru sites ($x=0.0625$)~[Fig.~\ref{RuO2fig5_4}\textcolor{blue}{(b)}], and two V atoms substituting for two Ru sites ($x=0.125$) [Fig.~\ref{RuO2fig5_4}\textcolor{blue}{(c)}].
Figures~\ref{RuO2fig5_4}\textcolor{blue}{(d)}, \ref{RuO2fig5_4}\textcolor{blue}{(e)}, and \ref{RuO2fig5_4}\textcolor{blue}{(f)} show the projected DOS onto the V-$d$, Ru-$d$, and O-$p$ states for pristine RuO$_2$, 6.25\% and 12.5\% V-substituted RuO$_2$, respectively.
Using the same $k$-point density as the primitive cell leads to artificially sharp peak features in the DOS of the $2\times2\times2$ supercell. 
To avoid such artificial peak features, DOS calculations for the periodic supercell with an ordered V arrangement were performed using a denser $8\times8\times12$ $k$-point mesh.
For $x=0.0625$, the Ru-$d$ DOS is almost unchanged from that of the pristine RuO$_2$, although the V-$d$ peaks are located slightly above the Fermi level.
For $x=0.125$, a shoulder of the V-$d$ peak begins to appear at the Fermi level, and the DOS increases considerably.
Some V-$d$ states are hybridized in the metallic band below $E_\mathrm{F}$.

\subsubsection{Tight-binding linear muffin-tin orbital (TB-LMTO) method}
The second approach is the TB-LMTO method based on DFT, using the coherent potential approximation (CPA).
It is implemented in the Korringa-Kohn-Rostoker (KKR) code \cite{PhysRevB.47.16532,ruban1999calculated}.
This method enables a description of random chemical substitution without explicitly constructing a large supercell, and it can directly treat concentration-dependent electronic structures of disordered alloys.

Figure~\ref{RuO2fig5_5} illustrates the structural model using the TB-LMTO method within the atomic sphere approximation (ASA). 
The chemical substitution of V atoms at Ru sites is treated using CPA, in which the Ru sublattice is substitutionally disordered, with each Ru site occupied by Ru and V with probabilities $1-x$ and $x$, respectively, while the O sublattice remains chemically ordered [Fig.~\ref{RuO2fig5_5}\textcolor{blue}{(a)}].
The exchange-correlation potential is treated within the GGA using the PBE functional form. 
The orbital basis and the structure-constant matrices are expanded up to the orbital quantum number $l=2$, i.e., $s$, $p$, and $d$ orbitals.
Since no additional Hubbard-$U$ correction is included, the present TB-LMTO-ASA+CPA calculations corresponds to $U = 0$.
As the ASA represents the crystal potential using overlapping atomic spheres that approximately fill the entire crystal volume, additional empty spheres are introduced into the interstitial regions of the rutile structure to achieve adequate space filling.
These empty spheres do not correspond to actual atoms but are
included as auxiliary sites in the TB-LMTO-ASA+CPA calculation. 
Table~\ref{tb4} lists the fractional coordinates and Wigner-Seitz radii used in the calculations, and Fig.~\ref{RuO2fig5_5}\textcolor{blue}{(b)} illustrates the corresponding atomic-sphere filling of the unit cell volume.
The lattice constants used are the same as those used in the PAW method.
The self-consistent field (SCF) calculations are performed using a $15\times15\times25$ $k$-point mesh in the full Brillouin zone (BZ) to obtain converged potentials.
The converged potentials are then used for DOS calculations with an $11\times11\times17$ $k$-point mesh in the full BZ and a smearing parameter of 1 mRy (13.6 meV).

Figure~\ref{RuO2fig6_6} shows the orbital-resolved DOS calculated using the TB-LMTO-ASA+CPA method for three compositions, namely, RuO$_2$, Ru$_{0.9}$V$_{0.1}$O$_2$, and Ru$_{0.8}$V$_{0.2}$O$_2$.
In RuO$_2$, the electronic states around the Fermi level predominantly originate from the Ru-$d$ orbitals.
For Ru$_{0.9}$V$_{0.1}$O$_2$, the peak of V-$d$ states emerges just above the Fermi level within 0.34 eV [Fig.~\ref{RuO2fig6_6}\textcolor{blue}{(b)}], and it reduces to 0.3 eV for Ru$_{0.8}$V$_{0.2}$O$_2$ [Fig.~\ref{RuO2fig6_6}\textcolor{blue}{(c)}].
With increasing substitution concentration up to 20\%, the shoulder of V-$d$ states appears at the Fermi level, leading to a considerable change in the total DOS.

\section{Conclusion}
In this study, we have grown Ru$_{0.9}$V$_{0.1}$O$_2$ single crystals using the sublimation transport method and investigated their physical properties.
Among the ten non-magnetic elements examined, only vanadium is uniformly substituted into the RuO$_2$ structure.
The average vanadium valence is nearly $+4$, while oxygen remains close to $-2$.
Upon 10\% V substitution, the room-temperature resistivity decreases by about 60\%, which is attributed to the substantial suppression of electron-phonon scattering.
The temperature-dependent paramagnetic susceptibility shows a slightly smaller positive coefficient up to room temperature in comparison with pristine RuO$_2$, with no indication of magnetic ordering.
Interestingly, the low-temperature upturn is noticeably weakened, indicating a reduction of localized paramagnetic impurity concentration.
These magnetic properties suggest that vanadium acts as a nonmagnetic substituent.
Electronic structure calculations using two distinct methods agree that the DOS at $E_\mathrm{F}$ does not change significantly up to 10\% V substitution, although the V-3$d$ states emerge just above $E_\mathrm{F}$.  
Thus, a higher level of V substitution may provide a viable route toward stabilizing an altermagnetic state.
These findings highlight nonmagnetic substitution in RuO$_2$ as a promising platform for exploring altermagnetic transitions and their experimental signatures.

\section*{acknowledgements}
We are grateful to A. S. Kasaliwal, G. Mattoni, T. Johnson, A. Ikeda for technical support. We acknowledge S. Souma, M. Sato, T. Osumi, and A. Eaton for useful discussions. This work was supported by the JSPS KAKENHI (JP22H01168, JP23K22439, JP25K17346, JP26K07012) and the JST Sakura Science Exchange Program (Grant No. Z2024L1015080).
We also acknowledge KU-Star program for its support.
S.P. acknowledges IIT Kanpur for financial support.
C.S. acknowledges research support from Anusandhan National Research Foundation (ANRF), Government of India (CRG-2022-005726, EEQ-2022-000883). H.M. acknowledges support from the Kyoto University Foundation.

\section{appendix}

\subsection{XPS fitting parameters}
The XPS spectra of Ru-$3p$, V-$2p$, and O-$1s$ were fitted using XPS PeakFit 4.1 software. During the fitting procedure, the intensity ratios of the Ru-$3p_{3/2}$/$3p_{1/2}$ and V-$2p_{3/2}$/$2p_{1/2}$ spin--orbit components were constrained to $2:1$ based on their respective spin--orbit degeneracies $(2j+1)$.
The resulting fitted parameters for RuO$_2$ and Ru$_{0.9}$V$_{0.1}$O$_2$ single crystals are summarized in Table~\ref{tb1b}.

\begin{table}[t]
    \setlength{\tabcolsep}{3pt}    
    \fontsize{8}{10}\selectfont
    \caption{XPS-peak fitting parameters for Ru-3$p$, V-2$p$, O-1$s$ spectra of RuO$_2$ and Ru$_{0.9}$V$_{0.1}$O$_{2}$ single crystals. \vspace{3pt}}
    \label{tb1b}
    \centering
    \begin{tabular}{cccccc} 
 	\hline 
    \hline 
     Core & Valency &B.E. & FWHM& Area\\
     line&  &(eV) & (eV)& (10$^{3}$~cts. eV/s)\\ 
    \hline
\textbf{RuO$_2$}&&&&\\   
    3$p_{3/2}$&Ru$^{4+}$& 462.8 & 3.97&5.39 \\
    3$p_{3/2}$&Ru$^{3+}$ & 466.4 & 4.20 & 1.07 \\
    3$p_{1/2}$&Ru$^{4+}$ & 485.0 & 4.27&2.69 \\
    3$p_{1/2}$&Ru$^{3+}$ & 488.7 & 4.10&0.54 \\  
    O$1s$&O$^{2-}$ & 529.5 & 1.36 &3.34 \\   
    O$1s$&O$^{1-}$ & 530.7 & 1.07 &1.0 \\  
    O$^\mathrm{Chem}$&H$_2$O & 532.8 & 2.22 &7.30 \\ 
    O$^\mathrm{Chem}$&C-O/C=O & 534.5 & 1.57 &1.85 \\  
    \hline
\textbf{Ru$_{0.9}$V$_{0.1}$O$_{2}$}&&&&\\
    3$p_{3/2}$&Ru$^{4+}$& 462.6 & 3.61&14.99 \\
    3$p_{3/2}$&Ru$^{3+}$ & 465.8 & 3.63&3.75 \\
    3$p_{1/2}$&Ru$^{4+}$ & 484.9 & 4.16& 7.5 \\
    3$p_{1/2}$&Ru$^{3+}$ & 488.1 & 4.0&1.87 \\  
    2$p_{3/2}$&V$^{3+}$& 515.0 & 1.78&2.37 \\
    2$p_{3/2}$&V$^{4+}$ & 516.6 & 2.09 & 7.92 \\
    2$p_{3/2}$&V$^{5+}$ & 517.3 & 1.50&1.97 \\
       Satellite &- & 518.9 & 1.50&0.98\\
    2$p_{1/2}$&V$^{3+}$ & 521.0 & 1.51& 1.18 \\
    2$p_{1/2}$&V$^{4+}$ & 523.0 &3.61&3.96 \\  
    2$p_{1/2}$&V$^{5+}$ & 524.4 &2.20&0.99 \\  
    O$1s$&O$^{2-}$ & 529.5 & 1.51 & 21.26 \\   
    O$1s$&O$^{1-}$ & 530.9 & 1.55 & 6.37 \\  
    O$^\mathrm{Chem}$&H$_2$O & 532.9 & 2.72 & 11.55 \\  
    O$^\mathrm{Chem}$&C-O/C=O & 534.7 & 1.90 & 1.48 \\  
    \hline
    \hline
 \end{tabular}
\end{table}

\begin{table}[H]
    \setlength{\tabcolsep}{4pt}
    \fontsize{8}{10}\selectfont
    \caption{Fractional coordinates and Wigner--Seitz radii used in the TB-LMTO-ASA+CPA calculations [as illustrated in Fig.~\ref{RuO2fig5_5}]. Here, “Em” denotes an empty sphere introduced solely for the ASA space-filling construction. The atomic and empty-sphere positions are specified in fractional coordinates with respect to the crystallographic unit cell. The Wigner-Seitz radius is expressed in unit of Bohr radius.\vspace{3pt}}
    \label{tb4}
    \centering
    \begin{tabular}{cccccc} 
 	\hline
    \hline 
X  &  Y  &       Z &    Atom  & Wigner-Seitz\\
  &    &        &     &  radius ($a_\mathrm{B}$)\\
    \hline 
0.00000 &  0.00000&   0.00000 &  Ru or V &  2.46080\\
0.50000  & 0.50000 &  0.50000 &  Ru or V&  2.46080   \\
0.69433   &0.30567  & 0.00000 &  O     &1.83575\\
0.80567  & 0.80567 &  0.50000&   O &    1.83575\\
0.19433 &  0.19433 &  0.50000  & O  &   1.83575\\
0.30567  & 0.69433  & 0.00000  & O  &   1.83575\\
0.50000  & 0.00000  & 0.25000  & Em &   1.71764\\
0.00000   &0.50000  & 0.25000 &  Em &   1.71764\\
0.00000   &0.50000 & -0.25000 &  Em  &  1.71764\\
0.50000  & 0.00000  &-0.25000 &  Em  &  1.71764\\
-0.31250 & -0.31250 & 0.00000&  Em   & 1.87021\\
-0.18750  & 0.18750 & 0.50000 & Em   & 1.87021\\
0.18750  &-0.18750  & 0.50000 & Em   & 1.87021\\
0.31250   &0.31250  & 0.00000  & Em   & 1.87021\\
    \hline
    \hline
 \end{tabular}
\end{table}

\subsection{Information of the TB-LMTO-ASA+CPA calculations}
\begin{figure}[H]
	\begin{center}
		\includegraphics[scale=0.5]{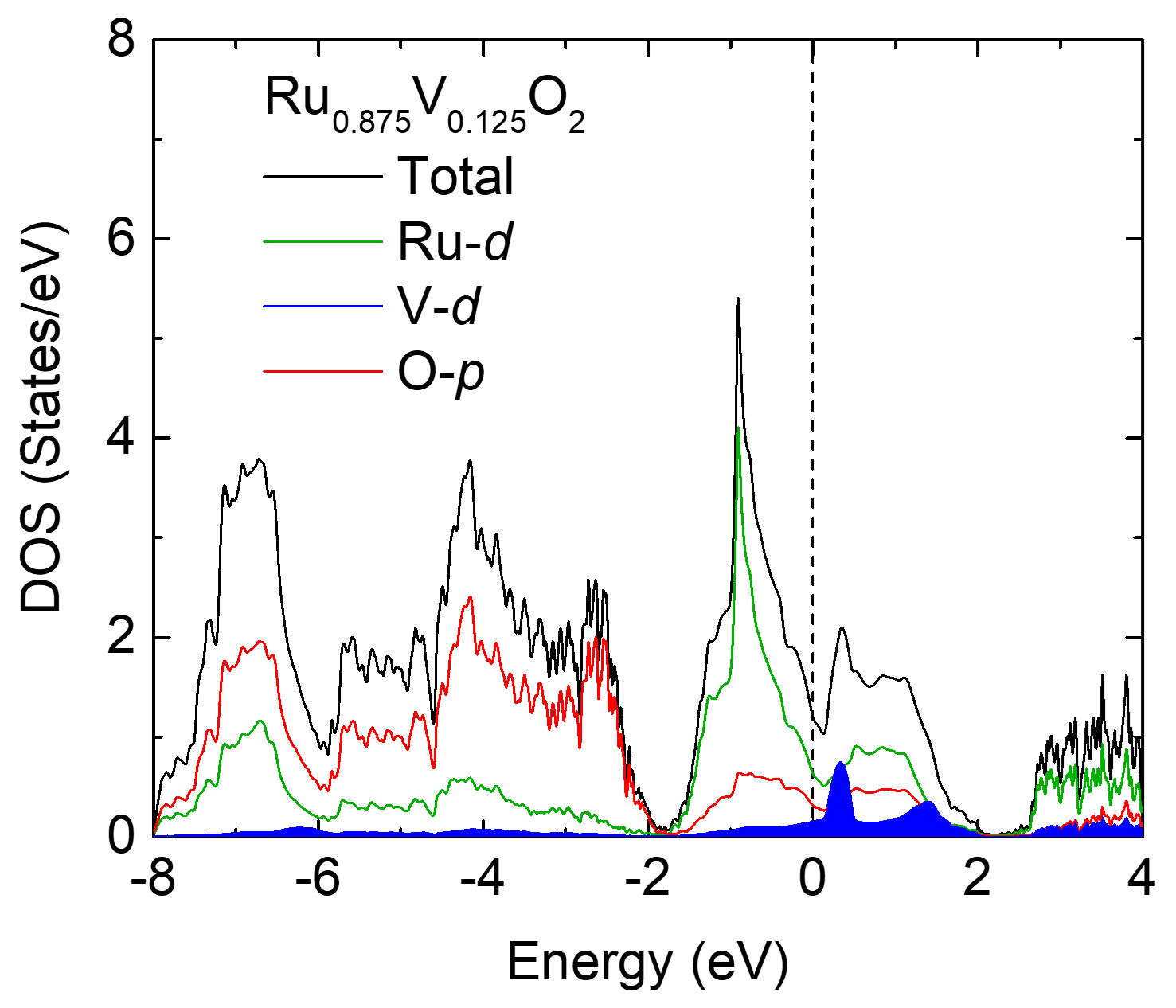}
		\caption{The orbital resolve electronic density of states (DOS) calculated using the TB-LMTO-ASA+CPA method for Ru$_{0.875}$V$_{0.125}$O$_2$.}
		\label{RuO2fig6_7}
	\end{center}
\end{figure}

The parameters used in the TB-LMTO-ASA+CPA calculations, corresponding to the illustrations in Fig.~\ref{RuO2fig5_5} are listed in Table~\ref{tb4}. 
We also performed the calculation based on this method for the vanadium concentration $x=0.125$ to make closer comparison with the PAW method shown in Fig.~\ref{RuO2fig5_4}\textcolor{blue}{(f)}.
The result shown in Fig.~\ref{RuO2fig6_7} is consistent with the PAW result and suggests that increasing vanadium concentration beyond 0.1 will steadily increase the DOS due to partial overlap with the V-3$d$ states just above the Fermi level. \\

\bibliography{ref}

@article{PhysRevB.47.16532,
  title = {Self-consistent linear-muffin-tin-orbitals coherent-potential technique for bulk and surface calculations: \ce{Cu}-\ce{Ni}, \ce{Ag}-\ce{Pd}, and \ce{Au}-\ce{Pt} random alloys},
  author = {Abrikosov, I. A. and Skriver, H. L.},
  journal = {Phys. Rev. B},
  volume = {47},
  issue = {24},
  pages = {16532--16541},
  numpages = {0},
  year = {1993},
  month = {Jun},
  publisher = {American Physical Society},
  doi = {10.1103/PhysRevB.47.16532}
}

@article{ruban1999calculated,
  title={Calculated surface segregation in transition metal alloys},
  author={Ruban, AV and Skriver, Hans Lomholt},
  journal={Computational Materials Science},
  volume={15},
  number={2},
  pages={119--143},
  year={1999},
  doi={10.1016/S0927-0256(99)00003-8},
  publisher={Elsevier}
}

@article{paul2025growth,
  title={Growth of ultra-clean oxide single crystals of the altermagnet candidate \ce{RuO2}},
  author={Paul, Shubhankar and Mattoni, Giordano and Matsuki, Hisakazu and Johnson, Thomas and Sow, Chanchal and Yonezawa, Shingo and Maeno, Yoshiteru},
  journal={Journal of Crystal Growth},
  volume={673 },
  pages={128405},
  year={2025},
  doi={https://doi.org/10.1016/j.jcrysgro.2025.128405},
  publisher={Elsevier}
}

@article{wu2025engineering,
  title={Engineering high-density microcrystalline boundary with \ce{V}-doped \ce{RuO2} for high-performance oxygen evolution in acid},
  author={Wu, Han and Fu, Zhanzhao and Chang, Jiangwei and Hu, Zhiang and Li, Jian and Wang, Siyang and Yu, Jingkun and Yong, Xue and Waterhouse, Geoffrey IN and Tang, Zhiyong and others},
  journal={Nature Communications},
  volume={16},
  number={1},
  pages={4482},
  year={2025},
  doi={https://doi.org/10.1038/s41467-025-59472-0},
  publisher={Nature Publishing Group UK London}
}

@article{kwok2000xps,
  title={XPS peak fitting program for WIN95/98 XPSPEAK version 4.1},
  author={Kwok, RWM},
  journal={Department of Chemistry, The Chinese University of Hong Kong},
  year={2000}
}

@article{he2025evidence,
  title={Evidence for single variant in altermagnetic \ce{RuO2} (101) thin films},
  author={He, Cong and Wen, Zhenchao and Okabayashi, Jun and Miura, Yoshio and Ma, Tianyi and Ohkubo, Tadakatsu and Seki, Takeshi and Sukegawa, Hiroaki and Mitani, Seiji},
  journal={Nature Communications},
  volume={16},
  number={1},
  pages={8235},
  year={2025},
  doi={https://doi.org/10.1038/s41467-025-63344-y},
  publisher={Nature Publishing Group UK London}
}

@article{smolyanyuk2025origin,
  title={Origin of the anomalous Hall effect in \ce{Cr}-doped \ce{RuO2}},
  author={Smolyanyuk, Andriy and {\v{S}}mejkal, Libor and Mazin, Igor I},
  journal={Physical Review B},
  volume={111},
  number={6},
  pages={064406},
  year={2025},
  doi={10.1103/PhysRevB.111.064406},
  publisher={APS}
}

@article{smolyanyuk2024fragility,
  title={Fragility of the magnetic order in the prototypical altermagnet \ce{RuO2}},
  author={Smolyanyuk, Andriy and Mazin, Igor I and Garcia-Gassull, Laura and Valent{\'\i}, Roser},
  journal={Physical Review B},
  volume={109},
  number={13},
  pages={134424},
  year={2024},
  doi={10.1103/PhysRevB.109.134424},
  publisher={APS}
}

@article{guo2024direct,
  title={Direct and inverse spin splitting effects in altermagnetic \ce{RuO2}},
  author={Guo, Yaqin and Zhang, Jing and Zhu, Zengtai and Jiang, Yuan-yuan and Jiang, Longxing and Wu, Chuangwen and Dong, Jing and Xu, Xing and He, Wenqing and He, Bin and others},
  journal={Advanced Science},
  volume={11},
  number={25},
  pages={2400967},
  year={2024},
  doi={doi.org/10.1002/advs.202400967},
  publisher={Wiley Online Library}
}

@article{wang2005structural,
  title={The structural and electronic properties of nanostructured \ce{Ce_{1-x-y}Zr_{x}Tb_{y}O2} ternary oxides: Unusual concentration of Tb3+ and metal-oxygen-metal interactions},
  author={Wang, Xianqin and Hanson, Jonathan C and Rodriguez, Jos{\'e} A and Belver, Carolina and Fern{\'a}ndez-Garc{\'\i}a, Marcos},
  journal={The Journal of chemical physics},
  volume={122},
  pages={154711},
  number={15},
  year={2005},
  doi = {10.1063/1.1883631},
  publisher={AIP Publishing}
}

@article{bat1995design,
  title={Design of \ce{RuO2}-based thermometers for the millikelvin temperature range},
  author={Bat'ko, I and Flachbart, K and Somora, M and Vanick{\`y}, D},
  journal={Cryogenics},
  volume={35},
  number={2},
  pages={105--108},
  year={1995},
doi = {https://doi.org/10.1016/0011-2275(95)92878-V},
  publisher={Elsevier}
}

@article{doi:10.1021/am4016445,
author = {Chun, Sung Hee and Choi, Hyun-A and Kang, Minkyung and Koh, Moonjee and Lee, Nam-Suk and Lee, Sang Cheol and Lee, Minyung and Lee, Youngmi and Lee, Chongmok and Kim, Myung Hwa},
title = {Highly Efficient Electrochemical Responses on Single Crystalline Ruthenium–Vanadium Mixed Metal Oxide Nanowires},
volume = {5},
number = {17},
pages = {8401-8406},
year = {2013},
doi = {10.1021/am4016445},
note ={PMID: 23977880},
journal = {ACS Applied Materials \& Interfaces}
}

@article{chun2013highly,
  title={Highly efficient electrochemical responses on single crystalline ruthenium--vanadium mixed metal oxide nanowires},
  author={Chun, Sung Hee and Choi, Hyun-A and Kang, Minkyung and Koh, Moonjee and Lee, Nam-Suk and Lee, Sang Cheol and Lee, Minyung and Lee, Youngmi and Lee, Chongmok and Kim, Myung Hwa},
  journal={ACS Applied Materials \& Interfaces},
  volume={5},
  number={17},
  pages={8401--8406},
  year={2013},
  doi={doi.org/10.1021/am4016445},
  publisher={ACS Publications}
}

@article{rodriguez1993recent,
  title={Recent advances in magnetic structure determination by neutron powder diffraction},
  author={Rodr{\'\i}guez-Carvajal, Juan},
  journal={Physica B: Condensed Matter},
  volume={192},
  number={1-2},
  pages={55--69},
 doi= {https://doi.org/10.1016/0921-4526(93)90108-I},
  year={1993},
  publisher={Elsevier}
}

@article{liu2026recent,
  title={Recent advances in the design of dual-doped \ce{RuO2} for efficient and stable acidic oxygen evolution},
  author={Liu, Jiqiang and Lv, Beibei and Ye, Yuheng and Deng, Yaoyao and Zhang, Chunyong and Bai, Jirong},
  journal={Nanoscale},
  volume={18},
  number={26},
  pages={13742--13754},
  year={2026},
  doi={doi.org/10.1039/d6nr01339b},
  publisher={The Royal Society of Chemistry}
}

@article{qian2025fragile,
  title={Fragile unconventional magnetism in \ce{RuO2} by proximity to Landau-Pomeranchuk instability},
  author={Qian, Zhuang and Yang, Yudi and Liu, Shi and Wu, Congjun},
  journal={Physical Review B},
  volume={111},
  number={17},
  pages={174425},
  year={2025},
  doi = {10.1103/PhysRevB.111.174425},
  publisher={APS}
}

@article{bose2022tilted,
  title={Tilted spin current generated by an antiferromagnet},
  author={Bose, Arnab and Schreiber, Nathaniel J and Jain, Rakshit and Shao, Ding-fu and Nair, Hari P and Sun, Jiaxin and Zhang, Xiyue S and Muller, David A and Tsymbal, Evgeny Y and Schlom, Darrell G and others},
  journal={Nat. Electron.},
  volume={5},
  pages={263--264},
  doi={https://doi.org/10.1038/s41928-022-00744-8},
  year={2022}
}

@article{morgan2015resolving,
  title={Resolving ruthenium: XPS studies of common ruthenium materials},
  author={Morgan, David J},
  journal={Surface and Interface Analysis},
  volume={47},
  number={11},
  pages={1072--1079},
  year={2015},
  doi={10.1002/sia.5852},
  publisher={Wiley Online Library}
}

@article{paul2025nonanalytic,
  title={Fermi-liquid behavior and characteristic temperature-dependent susceptibility in clean single crystal \ce{RuO2}},
  author={Paul, Shubhankar and Ikeda, Atsutoshi and Matsuki, Hisakazu and Mattoni, Giordano and Schmalian, J{\"o}rg and Yamauchi, Kunihiko and Sow, Chanchal and Yonezawa, Shingo and Maeno, Yoshiteru},
  journal={Physical Review B},
  volume={114},
  number={3},
  pages={034414},
  year={2026},
  doi={doi.org/10.1103/vy8x-q3l6},
  publisher={APS}
}

@article{paul2026CrSb,
  title={Thermodynamic and transport properties of high-quality single crystals of the altermagnet \ce{CrSb}},
  author={Paul, Shubhankar and Ikeda, Atsutoshi and Mattoni, Giordano and Yonezawa, Shingo and Sow, Chanchal},
  journal={Physical Review Materials},
  volume={10},
  number={6},
  pages={063402},
  year={2026},
  doi={doi.org/10.1103/hnjw-k826},
  publisher={APS}
}

@article{paul2026multi,
  title={Multi-probe detection of domain nucleation across the metal--insulator transition in \ce{VO2}},
  author={Paul, Shubhankar and Mattoni, Giordano and Ghosh, Amitava and Kesarwani, Pooja and Sahu, Dipak and Ahlawat, Monika and Verma, Amit and Govind Rao, Vishal and Sow, Chanchal and others},
  journal={Appl. Phys. Lett.},
  volume={128},
  number={5},
  pages = {2502226},
  year={2026},
  doi={10.1063/5.0291227},
  publisher={AIP Publishing}
}

@article{Over2000catalyst,
  title = {Atomic-Scale Structure and Catalytic Reactivity of the \ce{RuO2}(110) Surface},
  volume = {287},
  ISSN = {1095-9203},
  url = {http://dx.doi.org/10.1126/science.287.5457.1474},
  DOI = {10.1126/science.287.5457.1474},
  number = {5457},
  journal = {Science},
  publisher = {American Association for the Advancement of Science (AAAS)},
  author = {Over,  H. and Kim,  Y. D. and Seitsonen,  A. P. and Wendt,  S. and Lundgren,  E. and Schmid,  M. and Varga,  P. and Morgante,  A. and Ertl,  G.},
  year = {2000},
  month = feb,
  pages = {1474–1476}
}

@article{Ping2024catalyst,
  title = {Locking the lattice oxygen in \ce{RuO2} to stabilize highly active Ru sites in acidic water oxidation},
    author = {Ping,  Xinyu and Liu,  Yongduo and Zheng,  Lixia and Song,  Yang and Guo,  Lin and Chen,  Siguo and Wei,  Zidong},
  volume = {15},
  ISSN = {2041-1723},
  pages={2501},
  number = {1},
  journal = {Nature Communications},
  publisher = {Springer Science and Business Media LLC},
  year = {2024},
  DOI = {10.1038/s41467-024-46815-6},
  month = mar 
}

@article{wenzel2025fermi,
  title={Fermi-liquid behavior of nonaltermagnetic \ce{RuO2}},
  author={Wenzel, Maxim and Uykur, Ece and R{\"o}{\ss}ler, Sahana and Schmidt, Marcus and Janson, Oleg and Tiwari, Achyut and Dressel, Martin and Tsirlin, Alexander A},
  journal={Physical Review B},
  volume={111},
  number={4},
  pages={L041115},
  year={2025},
  doi= {https://doi.org/10.1103/PhysRevB.111.L041115},
  publisher={APS}
}

@article{berlijn2017itinerant,
  title={Itinerant antiferromagnetism in \ce{RuO2}},
  author={Berlijn, Tom and Snijders, Paul C and Delaire, O and Zhou, H-D and Maier, Thomas A and Cao, H-B and Chi, S-X and Matsuda, Masaaki and Wang, Yang and Koehler, Michael R and others},
  journal={Physical Review Letters},
  volume={118},
  number={7},
  pages={077201},
  year={2017},
  doi={https://doi.org/10.1103/PhysRevLett.118.077201},
  publisher={APS}
}

@article{mattheiss1976electronic,
  title={Electronic structure of \ce{RuO2}, \ce{OsO2}, and \ce{IrO2}},
  author={Mattheiss, LF},
  journal={Physical Review B},
  volume={13},
  number={6},
  pages={2433},
  year={1976},
  doi= {https://doi.org/10.1103/PhysRevB.13.2433},
  publisher={APS}
}

@article{mertig1986specific,
  title={Specific heat of Nb-Doped RuO2 single crystals},
  author={Mertig, M and Pompe, G and Hegenbarth, E},
  journal={Physica Status Solidi (b)},
  volume={135},
  number={1},
  pages={335--342},
  year={1986},
  doi={https://doi.org/10.1002/pssb.2221350133},
  publisher={Wiley Online Library}
}

@article{hiraishi2024nonmagnetic,
  title={Nonmagnetic Ground State in \ce{RuO2} Revealed by Muon Spin Rotation},
  author={Hiraishi, M and Okabe, H and Koda, A and Kadono, R and Muroi, T and Hirai, D and Hiroi, Z},
  journal={Physical Review Letters},
  volume={132},
  number={16},
  pages={166702},
  year={2024},
  doi={DOI: 10.1103/PhysRevLett.132.166702},
  publisher={APS}
}

@article{wu2025fermi,
  title={Fermi Surface of \ce{RuO2} Measured by Quantum Oscillations},
  author={Wu, Zheyu and Long, Mengmeng and Chen, Hanyi and Paul, Shubhankar and Matsuki, Hisakazu and Zheliuk, Oleksandr and Zeitler, Uli and Li, Gang and Zhou, Rui and Zhu, Zengwei and others},
  journal={Physical Review X},
  volume={15},
  number={3},
  pages={031044},
  year={2025},
doi={10.1103/5js8-2hj8},
 publisher={APS}
}

@article{naka2019spin,
  title={Spin current generation in organic antiferromagnets},
  author={Naka, Makoto and Hayami, Satoru and Kusunose, Hiroaki and Yanagi, Yuki and Motome, Yukitoshi and Seo, Hitoshi},
  journal={Nature communications},
  volume={10},
  number={1},
  pages={4305},
  year={2019},
  doi={https://doi.org/10.1038/s41467-019-12229-y},
  publisher={Nature Publishing Group UK London}
}

@article{hayami2019momentum,
  title={Momentum-dependent spin splitting by collinear antiferromagnetic ordering},
  author={Hayami, Satoru and Yanagi, Yuki and Kusunose, Hiroaki},
  journal={Journal of the Physical Society of Japan},
  volume={88},
  number={12},
  pages={123702},
  year={2019},
  doi={https://doi.org/10.7566/JPSJ.88.123702},
  publisher={The Physical Society of Japan}
}

@article{vsmejkal2020crystal,
  title={Crystal time-reversal symmetry breaking and spontaneous Hall effect in collinear antiferromagnets},
  author={{\v{S}}mejkal, Libor and Gonz{\'a}lez-Hern{\'a}ndez, Rafael and Jungwirth, Tom{\'a}{\v{s}} and Sinova, Jairo},
  journal={Science Advances},
  volume={6},
  number={23},
  pages={eaaz8809},
  year={2020},
  doi={10.1126/sciadv.aaz8809},
  publisher={American Association for the Advancement of Science}
}

@article{vsmejkal2022beyond,
  title={Beyond conventional ferromagnetism and antiferromagnetism: A phase with nonrelativistic spin and crystal rotation symmetry},
  author={{\v{S}}mejkal, Libor and Sinova, Jairo and Jungwirth, Tomas},
  journal={Physical Review X},
  volume={12},
  number={3},
  pages={031042},
  year={2022},
  doi={https://doi.org/10.1103/PhysRevX.12.031042},
  publisher={APS}
}

@article{vsmejkal2022emerging,
  title={Emerging research landscape of altermagnetism},
  author={{\v{S}}mejkal, Libor and Sinova, Jairo and Jungwirth, Tomas},
  journal={Physical Review X},
  volume={12},
  number={4},
  pages={040501},
  year={2022},
  doi={https://doi.org/10.1103/PhysRevX.12.040501},
  publisher={APS}
}

@article{karube2022observation,
  title={Observation of spin-splitter torque in collinear antiferromagnetic \ce{RuO2}},
  author={Karube, Shutaro and Tanaka, Takahiro and Sugawara, Daichi and Kadoguchi, Naohiro and Kohda, Makoto and Nitta, Junsaku},
  journal={Physical Review Letters},
  volume={129},
  number={13},
  pages={137201},
  year={2022},
  doi={https://doi.org/10.1103/PhysRevLett.129.137201},
  publisher={APS}
}

@article{Lin2024ARPES,
  author={Z. Lin and D. Chen and W. Lu and X. Liang and S. Feng and K. Yamagami and J. Osiecki and M. Leandersson and B. Thiagarajan and J. Liu and C. Felser and J. Ma},
  title={Observation of Giant Spin Splitting and d-wave Spin Texture in Room Temperature Altermagnet \ce{RuO2}},
  journal={preprint arXiv: 2402.04995},
  year={2024}
}

@article{ruf2021strain,
  title={Strain-stabilized superconductivity},
  author={Ruf, Jacob P and Paik, Hanjong and Schreiber, Nathaniel J and Nair, Hari P and Miao, Ludi and Kawasaki, Jason K and Nelson, Jocienne N and Faeth, Brendan D and Lee, Yonghun and Goodge, Berit H and others},
  journal={Nature Communications},
  volume={12},
  number={1},
  pages={59},
  year={2021},
  doi={https://doi.org/10.1038/s41467-020-20252-7},
  publisher={Nature Publishing Group UK London}
}

@article{uchida2020superconductivity,
  title={Superconductivity in uniquely strained \ce{RuO2} films},
  author={Uchida, Masaki and Nomoto, Takuya and Musashi, Maki and Arita, Ryotaro and Kawasaki, Masashi},
  journal={Physical Review Letters},
  volume={125},
  number={14},
  pages={147001},
  year={2020},
  doi={10.1103/PhysRevLett.125.147001},
  publisher={APS}
}

@article{osumi2025ARPES,
  title = {Spin-degenerate bulk bands and topological surface states associated with Dirac nodal lines in \ce{RuO2}},
  author = {Osumi, Takumi and Yamauchi, Kunihiko and Souma, Seigo and Paul, Shubhankar and Honma, Asuka and Nakayama, Kosuke and Ozawa, Kenichi and Kitamura, Miho and Horiba, Koji and Kumigashira, Hiroshi and Bigi, Chiara and Bertran, Fran\ifmmode \mbox{\c{c}}\else \c{c}\fi{}ois and Oguchi, Tamio and Takahashi, Takashi and Maeno, Yoshiteru and Sato, Takafumi},
  journal = {Physical Review B},
  volume = {113},
  issue = {8},
  pages = {085116},
  numpages = {18},
  year = {2026},
  month = {Feb},
  publisher = {American Physical Society},
  doi = {10.1103/wvs6-hqfv},
  url = {https://link.aps.org/doi/10.1103/wvs6-hqfv}
}

@article{zhu2019anomalous,
  title={Anomalous antiferromagnetism in metallic \ce{RuO2} determined by resonant x-ray scattering},
  author={Zhu, ZH and Strempfer, J and Rao, RR and Occhialini, CA and Pelliciari, J and Choi, Y and Kawaguchi, T and You, H and Mitchell, JF and Shao-Horn, Y and others},
  journal={Physical Review Letters},
  volume={122},
  number={1},
  pages={017202},
  year={2019},
  doi={10.1103/PhysRevLett.122.017202},
  publisher={APS}
}

@article{kessler2024absence,
  title={Absence of magnetic order in \ce{RuO2}: insights from $\mu$\ce{SR} spectroscopy and neutron diffraction},
  author={Ke{\ss}ler, Philipp and Garcia-Gassull, Laura and Suter, Andreas and Prokscha, Thomas and Salman, Zaher and Khalyavin, Dmitry and Manuel, Pascal and Orlandi, Fabio and Mazin, Igor I and Valent{\'\i}, Roser and others},
  journal={npj Spintronics},
  volume={2},
  number={1},
  pages={50},
  doi={https://doi.org/10.1038/s44306-024-00055-y},
  year={2024},
  publisher={Nature Publishing Group UK London}
}

@article{Wang-Kagawa2023NatComRuCrO2,
  title = {Emergent zero-field anomalous \ce{H}all effect in a reconstructed rutile antiferromagnetic metal},
  author = {Wang,  Meng and Tanaka,  Katsuhiro and Sakai,  Shiro and Wang,  Ziqian and Deng,  Ke and Lyu,  Yingjie and Li,  Cong and Tian,  Di and Shen,  Shengchun and Ogawa,  Naoki and Kanazawa,  Naoya and Yu,  Pu and Arita,  Ryotaro and Kagawa,  Fumitaka},
  DOI = {10.1038/s41467-023-43962-0},
  volume = {14},
  ISSN = {2041-1723},
  pages={8240},
  number = {1},
  publisher = {Springer Science and Business Media LLC},
  year = {2023},
  journal = {Nature Communications},
  month = dec 
}
\end{document}